\documentclass[10pt,twoside,BCOR7mm,DIV15,headinclude,footexclude,cleardoubleempty,idxtotoc]{scrartcl}

\usepackage[english]{babel}
\usepackage{graphicx}
\usepackage{hyperref}
\usepackage{scrpage2}
\usepackage{hyperref}
\usepackage{ifthen}

\makeatletter
\renewcommand{\@biblabel}[1]{}
\renewcommand{\@cite}[2]{%
{#1\ifthenelse{\boolean{@tempswa}}{,#2}{}}}
\makeatother

\hypersetup{breaklinks=true
,colorlinks=true,linkcolor=black,urlcolor=blue
,citecolor=black}

\ofoot{\thepage}
\ifoot{}

\setheadsepline{1pt}

\setkomafont{pagehead}{\normalfont\sffamily}
\setkomafont{pagenumber}{\normalfont\rmfamily}

\usepackage{booktabs}
\usepackage{amsmath}
\usepackage{amssymb}
\usepackage{multicol}
\usepackage{float}

\makeatletter
\newcommand{\listofcontributions}{\@starttoc{con}}

\newcommand{\l@contribution} {\@dottedtocline{1}{1.5em}{2.3em}}
\makeatother

\newenvironment{contribution}{
\setcounter{section}{0}
\setcounter{figure}{0}
\setcounter{table}{0}
\begin{flushleft}
{\em Clumping in Hot Star Winds \\
W.-R.\ Hamann, A.\ Feldmeier \& L.\ Oskinova, eds.\\
Potsdam: Univ.-Verl., 2007 \\
URN: http://nbn-resolving.de/urn:nbn:de:kobv:517-opus-13981
} 
\end{flushleft}
}{
\newpage
\lehead{}
\rohead{}
}

\begin{document}

\setlength{\baselineskip}{2.5ex}

\begin{contribution}
% EXAMPLE AND TEMPLATE FILE FOR PROCEEDINGS OF THE CLUMPING WORKSHOP.

% PLEASE REPLACE THE TEMPLATE TEXT BY YOUR OWN ARTICLE.

% NOTE THAT YOU MUST NOT PROCESS THIS FILE, BUT THE MASTER FILE:

% latex masterfile; dvips masterfile

% RUNNING AUTHOR: PUT AUTHOR NAMED HERE

\lehead{G. E.\ Romero, et al.}

% RUNNING TITLE; SHORTEN THE TITLE IF NECESSARY

% IN CASE OF A ONE-PAGE CONTRIBUTION (POSTER),

% SQUEEZE AUTHORS AND TITLE IN THIS LINE (Author: Title ...)

\rohead{Gamma-rays and the structure of stellar winds}

\begin{center}

% FULL TITLE HEADING

{\LARGE \bf Using gamma-rays to probe the clumped structure of stellar winds}\\

\medskip

% AUTHORS LIST

{\it\bf G. E. Romero$^{1,2}$, S. P. Owocki$^3$, A.T. Araudo$^{1,2}$, 
R. Townsend$^3$ \& P.\ Benaglia$^{1,2}$}\\

% AFFILIATIONS

{\it $^1$Instituto Argentino de Radioastronom\'{\i}a, C.C.5,

(1894) Villa Elisa, Buenos Aires, Argentina}\\

{\it $^2$Facultad de Ciencias Astron\'omicas y Geof\'{\i}sicas, Universidad Nacional de La Plata, 
Paseo del Bosque, 1900 La Plata, Argentina}\\

{\it $^3$Bartol Research Institute, University of Delaware, Newark, DE 19716, USA}

% ABSTRACT

\begin{abstract}

Gamma-rays can be produced by the interaction of a relativistic jet
and the matter of the stellar wind in the subclass of massive X-ray
binaries known as ``microquasars".  The relativistic jet is ejected
from the surroundings of the compact object and interacts with cold
protons from the stellar wind, producing pions that then quickly decay
into gamma-rays.  Since the resulting gamma-ray emissivity depends on
the target density, the detection of rapid variability in microquasars
with GLAST and the new generation of Cherenkov imaging arrays could be
used to probe the clumped structure of the stellar wind.
% in a variety of conditions.
In particular, we show here that the relative fluctuation in gamma
rays may scale with the square root of the ratio of porosity length to
binary separation, $\sqrt{h/a}$, implying for example a ca. 10\% variation
in gamma ray emission for a quite moderate porosity, $h/a \sim 0.01$.

\end{abstract}

\end{center}

% TEXT OF THE PAPER, TWO-COLUMN STYLE

\begin{multicols}{2}

\section{Introduction}

High-energy gamma-rays can be produced in accreting black holes with
relativistic jets, as shown by the recent detection of a high-energy
flare in the classical X-ray binary Cygnus X-1 (Albert et al.  2007).
In this type of system the primary object is a hot, massive star with
a strong stellar wind, and the secondary is usually an accreting black hole. 
Since the jet propagates through the wind and the radiation field of the
star, gamma-rays can be produced by inverse Compton scattering with
the radiation and/or pion-decays from inelastic proton-proton collisions
with the stellar wind.  
If the wind has a clumpy structure, then
jet-clump interactions can produce rapid flares of gamma-rays which,
if detected, could be used to probe the size, density, and velocity of
wind inhomogeneties.

\section{Gamma-ray emission from jet-clump interaction}

The basic model for hadronic gamma-ray production in a microquasar
with an homogeneous wind has been developed by Romero et al.  (2003).
The reader is referred to this paper for the basic formulae.  In the
case of a clumped wind, the situation is illustrated in Figure \ref{f1}.
The compact object is assumed to be in a circular orbit of radius $a$,
the clump crosses the jet forming an angle $\Psi$ with the center of
the star, and the viewing angle is $\theta$.

The spectral gamma-ray intensity (photons per unit of time per
unit of energy-band) produced by the injection of
relativistic protons in the clump is
\begin{equation}
I_{\gamma}(E_{\gamma},\theta)=\int_V n(\vec{r'})
q_{\gamma}(\vec{r'}) d^3\vec{r'}
\, , 
\label{I}
\end{equation}
where $V$ is the interaction volume, 
$n(\vec{r'})$ is the clump density, 
and $q_{\gamma}(\vec{r'})$ is the gamma-ray emissivity, 
which depends on the proton spectrum and the interaction cross section. The spectral energy distribution is given by $L^{\pi^0}_{\gamma}(E_{\gamma},\theta)=E_{\gamma}^2
I_{\gamma}(E_{\gamma},\theta)$.
Note that a change in the density of the target will be reflected in
a variation of the gamma-ray luminosity.  
Figure \ref{f2} shows a 3D-plot of the evolution of the spectrum during 
the transit of a clump through the jet.  
For this specific calculation we have assumed the clump flow follows
a typical $\beta$ velocity law at an angle $\Psi=5$ deg with the orbital
plane, with a viewing angle of 45 deg.  
The clump density follow Gaussian profile with peak at $\sim 10^{13}$ cm$^{-3}$ 
and a width of 0.01 $R_{\star}$.  
The primary star is assumed to be similar to the
primary in Cygnus X-1.  
The fraction of the accretion power that goes
to relativistic protons in the jet is $10^{-3}$.

From Fig.  \ref{f2} we see that a gamma-ray flare with a power-law
spectrum and a luminosity of $\sim 10^{35}$ erg s$^{-1}$ at 1 GeV is
produced by the interaction.  
The flare duration is set by the clump crossing time, which here is
a few hundred seconds.  
For clumps interacting at higher altitudes above the black hole, longer
timescales are possible, but the luminosity decreases since the jet
expands and hence the proton flux decreases.

%-----------One-column figure -----------------------------------

% Note that only the [H] option is allowed for placing 1-column figures!

\begin{figure}[H]
\begin{center}
\includegraphics[width=\columnwidth]{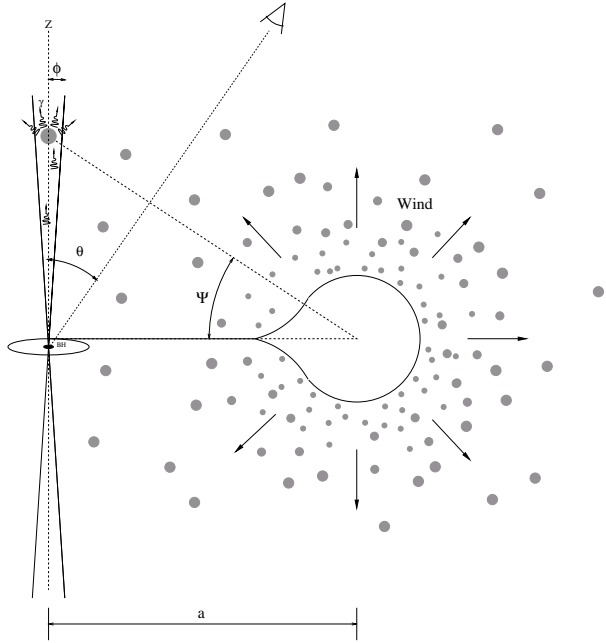}
\caption{
Sketch of a jet-clump interaction in a high-mass microquasar.}
\label{f1}
\end{center}
\end{figure}

%-----------One-column figure -----------------------------------

% Note that only the [H] option is allowed for placing 1-column figures!

\begin{figure}[H]
\begin{center}
\includegraphics[angle=270,width=\columnwidth]{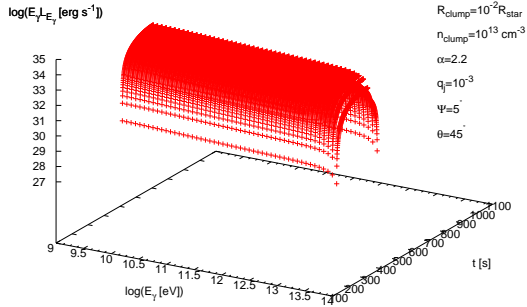}
\caption{
3D-plot with the light curve and the spectral energy distribution
resulting from a jet-clump interaction.  The main parameters adopted
are indicated in the figure.  See text for details.}
\label{f2}
\end{center}
\end{figure}

%-----------------------------------------------------------

% \section{Multi-clump interactions: a statistical approach}
\section{Porosity length scaling of gamma-ray fluctuation 
from multiple clumps}

Individual jet-clump interactions should be observable only as rare,
flaring events.  
But if the whole stellar wind is clumped, then integrated along the
beam there will be clump interactions occurring all the time,
leading to a flickering in the light curve, 
with the relative amplitude depending on the clump characteristics.
In particular, while the mean gamma-ray emission will depend on the mean number 
of clumps intersected, the relative fluctuation should (following
standard statistics) scale with the inverse square-root of this mean number.
But, as we now demonstrate, this mean number itself scales with the
same {\em porosity length} parameter that has been used, for example, 
by Owocki and Cohen (2006) to characterize the effect of wind clumps
on absorption of X-ray line emission.

Let us first consider the {\em mean} gamma-ray emission integrated along the
beam.
Assuming a narrow beam with constant total energy along its length
coordinate $z$,
the mean total gamma ray emission scales as
\begin{equation}
    \left < I_{\gamma} \right > = I_{b} \sigma \int_{0}^{\infty} n(z) \, dz
    \, ,
\label{iint}
\end{equation}
where $n(z)$ is the local mean wind density and $\sigma$ represents an
interaction cross-section for conversion of beam energy $I_{b}$ to gamma-rays.
For a steady wind with mass loss rate ${\dot M}$ and constant speed
$v$ emanating from a star at binary separation $a$ from the $z$-origin
at the black hole, the integral in Eq. (\ref{iint}) gives
\begin{equation}
    \left < I_{\gamma} \right > = {I_{b}  \sigma {\dot M_{\star}} \over 8 \mu v a }
    \, ,
\label{imean}
\end{equation}
where $\mu$ is the mean wind mass per interacting particle (e.g.,
protons), and $\dot M_{\star}$ is the star mass loss rate.

The fluctuation in this mean emission depends on the properties of the wind
clumps.
A simple model assumes a wind consisting entirely of clumps of
a  characteristic length $\ell$ and volume filling factor $f$, 
for which the mean-free-path for any ray through the clumps 
is given by the porosity length $h \equiv \ell/f$.
For a local interval along the beam $\Delta z$, the mean number of
clumps intersected is thus 
$\Delta N_{c} = \Delta z/h$, 
whereas the mean gamma-ray production is
% proportional to 
given by
\begin{equation}
    \Delta I_{\gamma}= I_{b} \sigma n \Delta z = I_{b} \sigma n \Delta N_{c} h.
\end{equation}
But by standard statistics for finite contributions from a discrete number
$\Delta N_{c}$, the {\em variance} of this emission is
\begin{equation}
    \left < \Delta I_{\gamma} ^{2} \right > 
    = { I_{b}^{2} \sigma^{2} n^{2} \Delta z^{2} \over \Delta N_{c} }
%     = { I_{b}^{2} \sigma^{2} n^{2} h^{2} \Delta N_{c} }
    = { I_{b}^{2} \sigma^{2} n^{2} h \Delta z } 
\, .
\end{equation}
The total variance is then just the integral that results from summing
these individual variances as one allows $\Delta z \rightarrow dz$,
\begin{equation}
    \delta I_{\gamma}^{2} 
    =  I_{b}^{2} \sigma^{2} \int_{0}^{\infty} n^{2} h d z 
\, .
\end{equation}
Taking the square-root of this yields an expression for the
relative rms fluctuation of intensity
\begin{equation}
{ \delta I_{\gamma} \over  \left < I_{\gamma}\right > }
= {\sqrt{  \int_{0}^{\infty}  n^{2} h \, dz } 
\over \int_{0}^{\infty} n \, dz
}
\, .
\end{equation}

As a simple example, for a wind with a constant velocity and constant porosity
length $h$, the relative variation is just
\begin{equation}
    { \delta I_{\gamma} \over  \left < I_{\gamma}\right > } = \sqrt{h/a} \, { 
\sqrt{\int_{0}^{\infty} dx/(1+x^{2})^{2}}
\over
\int_{0}^{\infty} dx/(1+x^{2})
}
= \sqrt{h/\pi a}
\, .
\end{equation}

On the other hand, for the Owocki \& Cohen (2006) uniform expansion model with 
$v \sim r$ and $h=h'r$, we find
$n \propto 1/r^{3}$ and thus,
\begin{equation}
    { \delta I_{\gamma} \over  \left < I_{\gamma}\right > } = \sqrt{h'} \, { 
\sqrt{\int_{0}^{\infty} dx/(1+x^{2})^{5/2}}
\over
\int_{0}^{\infty} dx/(1+x^2)^{3/2}
}
= \sqrt{2h'/3}
\, .
\end{equation}

Typically then, if $h\sim 0.03 a$, 
	\[	\frac{\delta I_{\gamma}}{\left < I_{\gamma} \right >}\approx 0.1.
\]
This implies an expected flickering at the level of 10\% for a
wind with such porosity parameters.  
The variability timescale will depend on the wind speed, the size of the
clumps, and the width of the jet, with a typical value of $\sim$ an hour.

\section{Prospects}

Fluctuations of 10\% in a source with a luminosity of $10^{34-35}$ erg
s$^{-1}$ on timescales of $\sim 10^{3}$ s could be detectable with
GLAST and CTA if the source is located at a few kpc.  This means that
gamma-ray astronomy can be used to probe the structure of stellar
winds through dedicated observations of microquasars with massive
donor stars.  If additional information on the wind can be obtained
through X-ray observations (Owocki \& Cohen 2006) and the source can
be detected with neutrino km$^3$-telescopes, we can also gain valuable
insights into the relativistic proton content of the jets.

\section{Acknowledgments}

G.E.R., A.T.A. and P.B. are supported by CONICET (PIP 5375) and the Argentine agency ANPCyT through Grant
PICT 03-13291 BID 1728/OC-AR.
S.P.O. acknowledges partial support of NSF grant 0507581 and NASA
Chandra grant TM7-8002X.

% REFERENCES IN ALPHABETICAL ORDER

\end{multicols}

\end{contribution}

%%-------------------------------------------------------

\end{document}